# Brownian motion and temperament of living cells

Roumen Tsekov[1] and Marga C. Lensen[2]
[1]Department of Physical Chemistry, University of Sofia, 1164 Sofia, Bulgaria
[2]Technische Universität Berlin, Institut für Chemie, 10623 Berlin, Germany

The migration of living cells obeys usually the laws of Brownian motion. While the latter is due to the thermal motion of surrounding matter, the cells locomotion is generally associated to their vitality. In the present paper the concept of cell temperament is introduced, being analogous to thermodynamic temperature and related to the cell entropy production. A heuristic expression for the diffusion coefficient of cell on structured surfaces is derived as well. The cell locomemory is also studied via the generalized Langevin equation.

Tissue cells (e.g. fibroblasts) are anchorage dependent, implying that they need an interface to adhere to. If a substrate with a sufficient rigidity is not available, those cells are not viable, even in the presence of extracellular matrix proteins. Successful initial adhesion is followed by cell spreading and eventually active migration, involving intricate mechano-transduction and biochemical signalling processes. Consequently, the cell migration is dependent on physical, chemical and mechanical cues, among others. This has inspired many researchers over the years to predict and control the cell migration, for example by administering increasing concentrations of nutrients to lure cells in a certain direction, which is denoted chemotaxis [1], or directing them to move up a gradient of substrate elasticity, a process called durotaxis [2, 3].

Controlling cell migration is a very important task in the biomedical field and in tissue engineering, since it determines for example the eventual integration of implants and plays an important role in cancer metastasis, where individual cells come loose from the tumour tissue and go out to settle at another suitable interface (e.g. in arteries). We have analysed the motility of single cells cultured on polymeric gel substrates, i.e. hydrogels, with variable rigidities. Our preliminary unpublished results show that both the mean square displacement and the average speed are larger on stiffer gels, while the mean square displacement scaled linearly with time, implying Brownian motion. Finally the persistence (angle of directional movement between subsequent steps) appeared larger on the softest gel, an observation that could however be affected inherently by the slower speed. Thus, it seems that cell motility and active migration are correlated to the substrate stiffness [2, 3], besides other factors such as the ones mentioned above. Nevertheless, from single cell observations it becomes clear that cells are individual entities and do not behave all identically, whereas the chemical or mechanical cues are supposedly homogeneous and act upon all cells similarly. We are interested to unravel other driving forces that dictate cellular behaviour and wish to find out more about cell individual characteristics. In a sense, we are searching for biophysical factors besides the well-known chemical and mechanical pa-rameters that dictate the cell behaviour in a certain situation. In addition, by looking at a statistically relevant number of single cells and analyse how their behaviour might deviate more or less from the average, we take into account the individuality of living cells.

Cell migration is usually described as Brownian motion [4-8] and non-Markovian effects are accounted for [9-14] as well. The general theory of Brownian motion is well developed in physics [15-17]. Starting from the Newtonian mechanics one can derive a generalized Langevin equation describing the stochastic dynamics of a Brownian particle [18]. An important result is the fluctuation-dissipation theorem, which relates the friction and stochastic forces in such a way that the system temperature remains constant. Although the theory of Brownian motion has been already applied to biological systems [19, 20], there are still open questions. For instance, the Brownian motion of particles is driven by thermal fluctuations in the surrounding, whereas the cells possess an inherent vital power. Hence, the thermal fluctuations are not the driving force of the cell migration. It is clear nowadays that cells are active Brownian particles [21-24]. The scope of the present paper is to study what drives cell migration and how to model memory effects in the Brownian motion of cells. We introduce the concept of cell temperament as an effective biophysical parameter driving the motion of living biological entities in analogy to the physical parameter temperature that dictates the movement of lifeless physical objects. We have explored a possibility to describe the cell locomemory via the Brownian self-similarity concept [25, 26].

Generally, the dynamics of physical systems is Hamiltonian. Although the cells are alive they can also be considered as physical objects [27] and can be described via a Liouville operator $i\hat{L}$. Hence, the evolution of the cell mass center $r(t)$ can be presented as

$$r(t) = \exp(i\hat{L}t)r \tag{1}$$

where $r \equiv r(0)$ is the initial point. Thus, the cell and its interactions with the environment are completely defined by the Liouville operator, which accounts for all physical and chemical effects. Though it is impossible to write a detailed expression for $i\hat{L}$, one can perform a general analysis without specification of the Liouville operator. Following the classical Mori-Zwanzig approach [16, 17] the exponential operator in Eq. (1) can be presented in an alternative way

$$\exp(i\hat{L}t) = \exp[(1-\hat{P})i\hat{L}t] - \int_0^t \exp(i\hat{L}s)\hat{P}i\hat{L}\exp[(1-\hat{P})i\hat{L}(t-s)]ds \tag{2}$$

where $\hat{P}$ is an arbitrary projection operator satisfying the general definition $\hat{P}^2 = \hat{P}$. A possible expression for it can be

$$\hat{P}r = v<vr>/<v^2> \qquad (3)$$

where $v = \dot{r}(0) = i\hat{L}r$ is the cell initial velocity. The projection operator (3) maps the coordinate space onto the velocity space of the cell. The brackets $<\cdot>$ indicate an empirical statistical average, which can differ from the usually employed equilibrium ensemble average. Thus, $<v^2>$ is the stationary non-equilibrium dispersion of cell velocity and reflects the cell vitality.

Using Eqs. (1) and (2) one can express the cell acceleration in the form

$$\ddot{r} = \exp[(1-\hat{P})i\hat{L}t]i\hat{L}v - \int_0^t \exp(i\hat{L}s)\hat{P}i\hat{L}\exp[(1-\hat{P})i\hat{L}(t-s)]i\hat{L}v ds \qquad (4)$$

Introducing now the stochastic Langevin force $f(t) \equiv m\exp[(1-\hat{P})i\hat{L}t]i\hat{L}v$, Eq. (4) acquires the form of a generalized Langevin equation [17]

$$m\ddot{r} + \frac{1}{m<v^2>}\int_0^t C_{ff}(t-s)\dot{r}(s)ds = f \qquad (5)$$

where $m$ is the cell mass. This equation describes a non-Markovian behavior and the stochastic force autocorrelation function $C_{ff} \equiv <f(t)f(s)>$ plays the role of a memory kernel.

The simplest description of the cell Brownian motion is a Markovian one, where the friction force is instantaneous and the stochastic cell dynamics obeys the ordinary Langevin equation [15]

$$m\ddot{r} + b\dot{r} = f \qquad (6)$$

where $b$ is the cell friction coefficient. If the cells do not perform active swimming [11] its friction coefficient can be estimated in liquids by the Stokes formula $b = 6\pi\eta R$, where $\eta$ is the liquid viscosity and $R$ is the cell radius [28, 29]. Comparing now Eqs. (5) and (6) yields an expression for the Langevin force autocorrelation function

$$C_{ff} = 2m<v^2>b\delta(t-s) \qquad (7)$$

which is a white noise with a constant spectral density. Equation (7) is the so-called second fluctuation-dissipation theorem. In the usual thermal Brownian motion the velocity dispersion of a Brownian particle is proportional to the system temperature $<v^2> = k_BT/m$ and the Langevin force autocorrelation function acquires the classical form $C_{ff} = 2k_BTb\delta(t-s)$. In the case of cell

Brownian motion $<v^2>$ is not determined by the temperature and it depends on the cell living power. For the sake of convenience we will introduce here an analogue of the thermodynamic temperature for living objects, $\theta \equiv m<v^2>$, which we call temperament. It is a measure of the cells living power reflecting in more intensive cell aspiration to move. The temperament $\theta$ is energy and is conceivingly proportional to the amount of energy stored as ATP in cells.

In the general case the cell migration can be also affected by external forces $F$, which can cause for instance chemotaxis [1], durotaxis [2, 3], etc. In this case the existing gradient of the surface properties forces the cells to migrate along to. Of course, due to the random movement, the cells experience in fact a directed active Brownian motion. The force balance of the cell in this case reads

$$m\ddot{r} + b\dot{r} = f + F \tag{8}$$

Taking an average value of this equation and remembering that the fluctuation force possesses zero mean value one yields an equation for the mean displacement $<r(t)>$ of the cell

$$m<\ddot{r}> + b<\dot{r}> = F \tag{9}$$

Usually the friction force is much larger than the cell acceleration and in this case Eq. (9) reduces to $<\dot{r}> = F/b$. If the external force is constant in time one can derive the following relation for the cell mean displacement $<r> = Ft/b$, which depends linearly on time. Hence, if several cells are monitored one can calculate their average displacement and plotting it versus time one could obtain the ratio $F/b$ from the slope of the linear fit. Thus if the external force is known one is able to determine experimentally the friction coefficient $b$, which is an interesting characteristic of the cells motion. It depends on how the cells interact with the surrounding. In practice, apart from the gravitational force in sedimentation, the force $F$ is unknown in chemotaxis and durotaxis. For this reason, one uses typically the decay of the velocity autocorrelation function of the cell migration to extract the friction coefficient $b$.

Subtracting now Eq. (9) from Eq. (8) yields an equation for the cell position fluctuations

$$m(\ddot{r} - <\ddot{r}>) + b(\dot{r} - <\dot{r}>) = f \tag{10}$$

This equation corresponds to pure Brownian motion and for this reason the dispersion of the cell position is given by $<r^2> - <r>^2 = 2Dt$. Hence, plotting this linear relation from experimental measurements one is able to obtain the cell diffusion constant $D$ as well. According to Eq. (7) it is given by the ratio of the cell temperament and friction coefficient, $D = \theta/b$. Since the friction coefficient $b$ can be calculated independently from the mean displacement, the cell temperament $\theta = bD$ can be estimated from these two experimentally measured coefficients [8]. Thus, one could calculate this very interesting characteristic of the cell activity. It is exciting to compare the temperament of different types of cells.

The cell temperament is related to the entropy production $\dot{S}$ in cells which according to the non-equilibrium thermodynamics depends on thermodynamic flows $\{\dot{x}_k\}$ and forces $\{f_k\}$

$$\dot{S} = \sum_k (\partial S / \partial x_k) \dot{x}_k = \sum_k f_k \dot{x}_k = \sum_k \sum_n L_{kn} \dot{x}_n \dot{x}_k \qquad (11)$$

The last expression is valid for the linear non-equilibrium thermodynamics, where $\{L_{kn}\}$ is the symmetric matrix of kinetic coefficients. In the case of cell motion, one of the generalized flows is obviously the cell velocity $v \equiv \dot{x}_1$. The coefficient $L_{11} = b/T$ equals to the friction coefficient divided by the temperature, while since the translation is a vector process $L_{1n \neq 1} = 0$ due to the Curie principle. Thus, the temperament $\theta = m < \dot{x}_1^2 >$ can be expressed as follows

$$\theta = T(<\dot{S}> - <\dot{S}_{-1}>)m/b \qquad (12)$$

where $<\dot{S}_{-1}> = \sum_{k \neq 1} \sum_{n \neq 1} L_{kn} <\dot{x}_n \dot{x}_k>$ is the average entropy production due to all other activities of the cell excluding the cell migration. Equation (12) suggests proportionality between the temperament and temperature. It could be used also for calculation of the cell entropy production, which is further related to specific biophysical processes. If the cell is not alive $<x_k f_n> = k_B \delta_{kn}$ according to the equipartition theorem and the cell will produce during the momentum relaxation time $m/b$ kinetic entropy equal to the Boltzmann constant $k_B$. Hence, in this case the temperament will coincide with the thermodynamic temperature $\theta = k_B T$ and the cell will move as a usual Brownian particle. If the cell is alive it will produce much more kinetic entropy to compensate the active entropy flow from the environment and keep the stationary state of life. Hence, the temperament will be much higher than $k_B T$, which will reflect in more intensive Brownian motion and higher diffusion constant $D$ [30]. Indeed, the corresponding effective temperature, estimated by the Stokes-Einstein formula $\theta = 6\pi\eta R D$ from experimental values of the cell diffusion constant, is several orders of magnitude higher than the thermodynamic one [28, 29].

An interesting aspect of the cell migration is the Brownian motion of cells attached on structured surfaces. In this case the cells experience a periodic potential $U$, which modulates their motion. The corresponding Langevin equation reads

$$m\ddot{r} + b\dot{r} = f - \partial_r U \qquad (13)$$

Usually the friction on a surface is much stronger than in the bulk and, for this reason, the inertial effects in Eq. (13) can be neglected. Thus, following Eq. (7) the probability density $\rho(r,t)$ to find the cell at the point $r$ at the moment $t$ obeys the Smoluchowski equation

$$\partial_t \rho = \partial_r \cdot (\rho \partial_r U + \theta \partial_r \rho)/b \qquad (14)$$

where the temperament plays the role of an effective temperature. The term $\theta \rho$ represents the living cell osmotic pressure and, hence, the cell temperament $\theta$ can be conveniently measured by osmotic experiments as well. In this respect an interesting question arises here: what is the osmotic pressure of fishes in an aquarium? Hence, the temperament can be attributed also to animals, people, etc. The stationary distribution, provided by Eq. (14), is a Boltzmann-like probability density $\rho_{st} \sim \exp(-U/\theta)$. In the case of gravity with $U = mgz$ Eq. (14) describes the barometric distribution of cells with stationary probability density $\rho_{st} = (mg/\theta)\exp(-mgz/\theta)$, while small vibrations of an adsorbed cell are described by a harmonic potential $U = m\omega_0^2 x^2/2$ and probability density $\rho_{st} = \sqrt{m\omega_0^2/2\pi\theta}\exp(-m\omega_0^2 x^2/2\theta)$.

In the field of a periodic potential $U$ the cell undergoes a continuous Brownian motion with an effective diffusion coefficient calculated from Eq. (14) [31, 32]

$$D_{eff} = D/<\exp(-U/\theta)><\exp(U/\theta)> \qquad (15)$$

where the brackets $<\cdot>$ indicate spatial geometric average along the surface. For instance, in the case of a cosine potential $U = A\cos(qx)$ the explicit calculation is possible and the effective diffusion coefficient acquires the form

$$D_{eff} = D/I_0^2(A/\theta) \qquad (16)$$

where $I_0$ is the modified Bessel function of first kind and zero order. If the height of the potential barriers $2A$ is much smaller than the cell temperament $\theta$ the effective diffusion coefficient reduces to that one on a non-structures surface $D$. In the opposite case of strong barriers Eq. (16) approximates well by the Arrhenius-like dependence

$$D_{eff} = 2\pi(A/b)\exp(-2A/\theta) \qquad (17)$$

Hence, the temperament is very essential for description of the cell motion in potential landscapes, which is the usual case in practice. The direct parallel between the temperament and temperature allows employment of many well-known results from the statistical mechanics for description of phenomena related to living objects. A similar approach for accounting of the effective quantum temperature is used in chemical kinetics and catalysis [33].

From Eq. (13) one can derive in a standard way the Klein-Kramers equation describing the evolution of the probability density $W(v,r,t)$ in the cell phase-space

$$\partial_t W + v \cdot \partial_r W - \partial_r U \cdot \partial_v W / m = b \partial_v \cdot (vW + \theta \partial_v W / m) / m \qquad (18)$$

It provides the Maxwell-Boltzmann-like distribution $W_{st} \sim \exp[-(mv^2/2+U)/\theta]$ as the equilibrium solution. The Smoluchowski equation (14) can be derived from Eq. (18) in the case of large friction. In the case of a free cell Eq. (18) can be integrated directly along the cell coordinate $r$ to obtain the Fokker-Planck equation for the probability density $w(v,t)$ in the cell velocity space

$$\partial_t w = b \partial_v \cdot (vw + \theta \partial_v w / m) / m \qquad (19)$$

This equation shows that the cell velocity is an Ornstein-Uhlenbeck process.

Due to the active locomotion of the cells the fluctuation Langevin force $f$ is strongly related to the cell vitality. How it was demonstrated above the simple white noise model is very useful but it does not take into account the memory effects, which are important in the living world. According to Eq. (5) the memory function of the Brownian motion is determined by the Langevin force autocorrelation function, which is a manifestation of the fluctuation-dissipation theorem. Since the Langevin force is not correlated to the initial cell velocity, $<f(t)v>=0$, it is straightforward to derive from Eq. (5) an integro-differential equation for the cell velocity autocorrelation function $C_{vv}(t) \equiv <v(t)v>$

$$m\theta \dot{C}_{vv}(t) + \int_0^t C_{ff}(t-s) C_{vv}(s) ds = 0 \qquad (20)$$

Using the standard Laplace transformation this equation can be transformed to

$$\tilde{C}_{vv}(p)/D = 1/[mp/b + \tilde{C}_{ff}(p)/b\theta] \qquad (21)$$

where $p$ is the Laplace transform variable. As is seen, Eq. (21) relates the memory function to the Laplace spectral density of the cell velocity fluctuations. A useful idea to close the problem is to supply an additional relationship between these two quantities, which is the essence of the concept of Brownian self-similarity [25].

In a previous paper [34] we proposed similarity between the Laplace spectral densities of the Langevin force and Brownian particle velocity as a model of hydrodynamic fluctuations. Ac-cording to this concept the form of the Langevin force autocorrelation function spectral density is the same as that of $\tilde{C}_{vv}$ in Eq. (21)

$$\tilde{C}_{ff}(p)/b\theta = 1/[p\tau + \tilde{C}_{ff}(p)/b\theta] \qquad (22)$$

Here the new parameter $\tau$ is the correlation time of $f$ and Eq. (22) provides the standard white noise expression $\tilde{C}_{ff}(p) = b\theta$ if $\tau = 0$. In general, the solution of Eq. (22) reads

$$\tilde{C}_{ff} = b\theta[\sqrt{1+(p\tau/2)^2} - p\tau/2] \qquad C_{ff} = b\theta J_1(2t/\tau)/t \qquad (23)$$

where $J_1$ is the Bessel function of first kind and first order. The plot of this Langevin force autocorrelation function in Fig. 1 exhibits an oscillatory behavior corresponding to a sequence of correlations and anticorrelations. Another interesting property of the Langevin force autocorrelation function (23) is the long-time tail, where the amplitude of $C_{ff}$ decays in time as $1/t^{3/2}$. The Langevin force possesses also a finite dispersion $C_{ff}(0) = b\theta/\tau$.

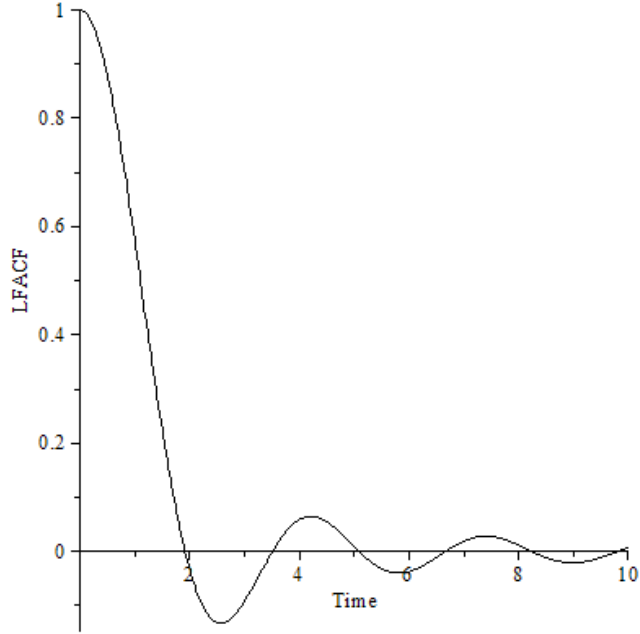

**Fig. 1** The dependence of $C_{ff}(t)/C_{ff}(0)$ on the dimensionless time $t/\tau$ according to Eq. (23).

Substituting Eq. (23) in Eq. (21) leads to an expression for the Laplace spectral density of the cell velocity autocorrelation function

$$\tilde{C}_{vv}(p) = D/[mp/b + \sqrt{1+(p\tau/2)^2} - p\tau/2] \qquad (24)$$

This is the so-called Rubin model [35] known from the physics of chains of oscillators. Unfortunately, it is impossible to invert the Laplace image (24) in general but there are some particular cases, where analytical expressions are obtained. If the cell relaxation time $m/b$ is much larger that the correlation time of the Langevin force $\tau$ Eq. (24) reduces to the standard exponentially decaying function $C_{vv} = (\theta/m)\exp(-bt/m)$. If these two time constants are equal, the velocity autocorrelation function $C_{vv} = DJ_1(2bt/m)/t$ coincides with the Langevin force autocorrelation function (23). This case corresponds to Brownian particles driven by similar objects [25]. From this perspective cells, which dissipate energy mainly insides, should possesses velocity autocorrelation function $C_{vv}$ like that in Fig. 1. The remaining of the heat into the cells is a requirement to keep the cell temperament constant in the case of absent external energy flow. If $\tau = 2m/b$ the inverse Laplace image of Eq. (24) is another oscillatory-decaying function $C_{vv} = (\theta/m)J_0(bt/m)$, where $J_0$ is the Bessel function of first kind and zero order. Finally, if the correlation time of the Langevin force is infinite, the cell velocity is completely correlated at any time, $C_{vv} = \theta/m$.

The Brownian motion has been originally discovered by colloidal particles but later on it is realized that BM is a universal movement of matter. For this reason our application to the cell motion started from the first principles to convince the reader that cells should also follow the Brownian motion dynamics. Therefore, there is no room for any doubt in the applicability of the Brownian motion model to living cells and the only new specific parameter introduced is the cell velocity dispersion being directly proportional to the temperament. The observed deviations from the Einstein law are due, however, to the Gaussian white noise model and can be resolved by proper modelling of the memory function in Eq. (5) and such an example is given in the previous section of the paper. Even in the exact generalized Langevin equation (5) the problem for the value of the temperament is central. It is the cell property and the effect of the environment is to contribute a negligible addition $k_B T$ to the temperament $\theta$. The more essential in a structured environment is the modulation of the friction coefficient $b$ and the effective periodic potential $U$ via surface topography, stiffness, etc. These effects are very interesting both for theoretical modelling or experimental measurement but they are out of the scope of the present paper. For illustration we have mentioned in the paper the simplest case of spherical cells, where $b$ is given by the Stokes law, and estimated the corresponding values of temperament from already published data.